# Dagger: Towards Efficient RPCs in Cloud Microservices with Near-Memory Reconfigurable NICs

Nikita Lazarev, Neil Adit, Shaojie Xiang, Zhiru Zhang, and Christina Delimitrou

**Abstract**— Cloud applications are increasingly relying on hundreds of loosely-coupled microservices to complete user requests that meet an application's end-to-end QoS requirements. Communication time between services accounts for a large fraction of the end-to-end latency and can introduce performance unpredictability and QoS violations. This work presents our early work on *Dagger*, a hardware acceleration platform for networking, designed specifically with the unique qualities of microservices in mind. The Dagger architecture relies on an FPGA-based NIC, closely coupled with the processor over a configurable memory interconnect, designed to offload and accelerate RPC stacks. Unlike the traditional cloud systems that use PCIe links as the NIC I/O interface, we leverage memory-interconnected FPGAs as networking devices to provide the efficiency, transparency, and programmability needed for fine-grained microservices. We show that this considerably improves CPU utilization and performance for cloud RPCs.

## 1 Introduction

The past few years have seen a major shift in the way cloud applications are designed, from traditional monolithic architectures to microservices. While microservices improve error isolation and accelerate development and deployment iterations, they also introduce non-negligible communication overheads in terms of networking [8], [19]. Remote Procedure Calls (RPC) are one of the most commonly-used communication techniques in microservices. There is a variety of commercially available RPC frameworks; however, since these frameworks were not designed specifically with microservices in mind, they do not address their unique resource requirements.

As the demand for high-bandwidth and low-latency networking in the cloud continues to grow, research from both industry and academia has offered numerous proposals for efficient datacenter network designs. One set of proposals is based on user-space dataplane architectures pushing a larger fraction of the networking stack from kernel to user space [4], [10]. While this approach eliminates the overhead of the user/kernel boundary crossing, it still implements the entire networking stack in software, therefore consuming processor resources and being subject to the generality overheads of CPU-based execution.

Another line of work proposes offloading network processing to specialized adapters, and leverages RDMA to implement high level communication primitives, such as RPCs [5], [12]. While prior work has demonstrated the efficiency benefits of RDMA compared to traditional networking, since commercially available network adapters are typically viewed by the processor as peripheral devices over PCIe, such RDMA-based architectures still suffer from unnecessary overheads from the interconnect's messaging model [7], [11], [13]. Moreover, the proposed solutions do not offload the entire networking stack to hardware, leaving computationally-intensive application layers (e.g., RPC data transformation, (de)encryption, (de)compression, etc.) to the host processor.

To address the aforementioned issues, prior work has proposed closely integrating NICs with CPUs. NetDIMM [2], for example, suggests embedding network interfaces into DIMM memory hardware. While the idea has merit, implementing it in silicon and performing an end-to-end evaluation requires taping-out custom memory hardware, which is problematic given the frequency with which cloud services change. soNUMA [14] offers a more practical solution by scaling out coherent NUMA interconnects at the datacenter level, therefore providing fast and efficient networking. NeBuLa [21] extends soNUMA and discusses the implementation of a hardware-based RPC stack over scaled NUMA interconnects, and further proposes to directly deliver the RPC payload all the way to the L1 cache. The NeBuLa architecture offers dramatic network queueing reduction and improved throughput and latency for RPCs. However, NeBuLa requires changing the processor's memory system and fabricating dedicated silicon, a non-trivial undertaking when targeting datacenter-scale deployments. The same applies for Optimus Prime [16], which proposes offloading RPC data transformations to a closely-coupled integrated ASIC accelerator.

In this work, we present Dagger, the first end-to-end FPGA-based reconfigurable RPC stack integrated with the processor over a NUMA-like memory interconnect. In contrast to soNUMA [14], Dagger does not rely on scaling-out complex memory interconnects and implementation of the RPC stack in custom silicon, but rather follows the standard design architecture of networking systems where the machines in a cluster are connected over commodity Ethernet. Similarly to NeBula and Optimus Prime, we leverage memory interconnects, but only as the interface between the processor and the NIC as an Ethernet device. We use FPGAs as the physical medium for Dagger to (1) make the implementation of the near-memory RPC hardware feasible today without the need for taping-out custom hardware, and to (2) make the NIC reconfigurable, so the networking stack can dynamically adjust to the running workload. The latter is especially important for microservices, which are by design very diverse and frequently updated [8], [18]. While several prior proposals have discussed the potential of programmable NICs [3], [6], [17], they all rely on commodity PCIe technology for the interconnect, and are not focused on hardware-based RPC systems, therefore, leaving the application part of the RPC protocol stack running on the host processor.

We characterize the unique properties of microservice network footprints using the Social Network and Media Service applications from the DeathStarBench suite [8], and prototype Dagger on the Intel Broadwell integrated CPU/FPGA architecture available in today's cloud systems. We demonstrate in practice that offloading networking to a near-memory FPGA significantly increases per-core RPC throughput for the small requests common in microservices up to $2.4-3.8\times$, compared to both specialized hardware platforms [12] and optimized software protocols [4], [10]. Our solution yields single-thread throughput of 12.4 Mrps; it scales up to 42 Mrps with only four threads on two CPU cores, and provides state-of-the-art $\mu$s-scale end-to-end latency.

## 2 Network Characteristics in Microservices

Microservices have distinct network requirements and traffic patterns, compared to monolithic applications and traditional distributed systems. First, every user request in microservices is propagated through a large graph of tiers, with per-node processing and communication delays being accumulated in the end-to-end latency. As a result, the Quality of Service (QoS), which is usually defined in terms of tail latency under a certain load in Queries per Second (QPS), critically depends on the performance of every communication channel between each pair of microservices on the call path. Hence, even a small latency increase in the networking stack translates to considerable increases in end-to-end latency, as shown in Figure 1 which plots the end-to-end fractions of networking and application time (including queueing) with respect to load.

Second, even though RPC request and response payloads in typical datacenter applications are already relatively small, ranging from hundreds of bytes to few kBytes [9], [22], in microservices that number is even smaller, as shown in Figure 2 for the Social Network and Media Service from DeathStarBench [8].

- *All authors are with the School of ECE, Cornell University, Ithaca, NY. E-mail: {nl524, na469, sx233, zhiruz, delimitrou}@cornell.edu*

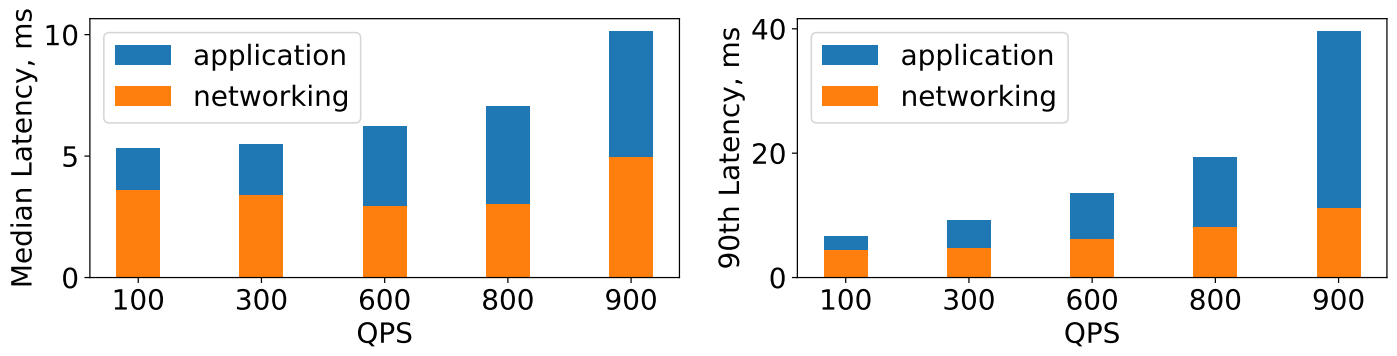

Fig. 1: Networking as a fraction of median (left) and tail (right) latency.

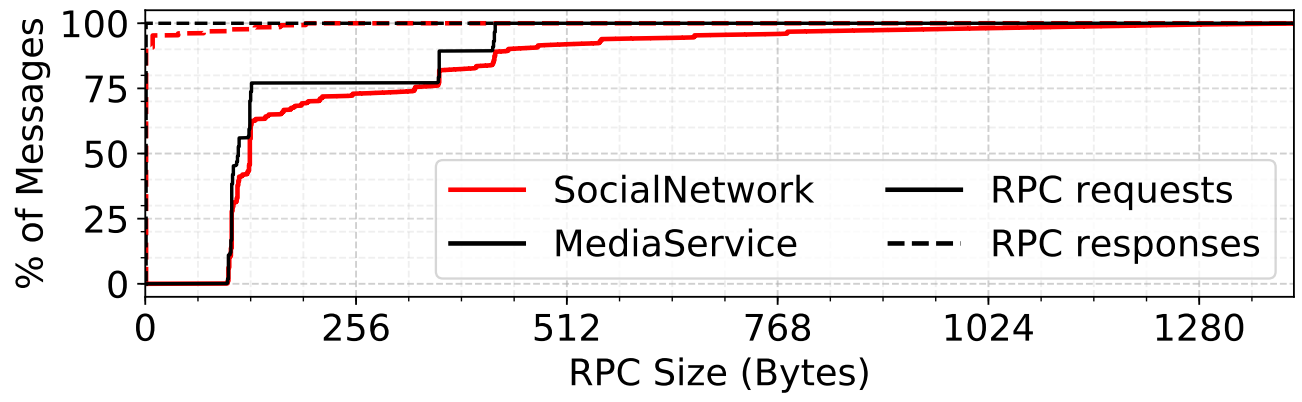

Fig. 2: Distribution of RPC sizes across microservices in [8].

Figure 2 shows that more than 70% of RPC requests are smaller than 256B, and almost all requests are within 1280B. Responses are even smaller: nearly 100% of messages are less than 256B with 95% of messages being under 64B. These tiny messages introduce high pressure on networking stacks at all levels. In fact, previous work has shown that commodity networking systems cannot efficiently handle this traffic, due to high per-packet overheads [4], [10].

Finally, microservices are by design very diverse in terms of their design patterns and performance requirements [18]. In particular, there is a rich variety of thread models [20], network queueing architectures [21], and strategies of mapping microservices to the available hardware resources. Performance requirements also vary, with some microservices being latency-critical, while others are treated as batch. Commodity networking systems do not necessarily provide the most efficient solution for a particular application class. This has caused programmable networking systems to become more popular [3], [15]. Such systems allow flexibly adjusting networking primitives depending on the currently-running applications. Even so, all today's programmable NICs are based on a fixed CPU-NIC interface.

## 3 Dagger: A Near-Memory Reconfigurable NIC for Interactive Microservices

### 3.1 Integrating NICs over Memory Interconnects

PCIe is the current de facto standard interface between processors and NICs or accelerators. Unfortunately, the PCIe protocol has limited functionality, it requires multiple transactions and explicit memory synchronization to send data chunks to the device, which increases the per-packet overheads. The regular way to send data over PCIe is by using DMA transfers together with expensive notification (initiation) transactions explicitly issued by the processor as memory-mapped I/O (MMIO) requests (also known as doorbells [11]). Various optimizations can be applied to improve the efficiency of doorbells [7], [11], although none of them can completely avoid the expensive MMIOs.

The fundamental disadvantage of the traditional PCIe model when it comes to interactive microservices is caused by following the Producer-Consumer ordering models common in the majority of peripheral devices, which work well on data streams and large continuous objects [1]. However, microservice RPC flows do not conform to such patterns: (1) their small fine-grained RPCs are normally independent of each other; (2) they do not necessarily arrive in a streaming fashion; and (3) when optimizing for low latency, they cannot be efficiently batched to form large data streams. When using production-level RPC frameworks, such as Apache Thrift, RPC requests can be discontiguous in memory and may contain references to nested data structures, requiring expensive and non-zero-copying serialization [16] on the CPU for commodity doorbells.

The main advantage of using memory interconnects as the CPU-NIC interface is that the data transfer can be handled entirely in hardware without the need for explicit notifications from the processor. In the best-case scenario, the only operation the processor should do in the critical path is preparing an RPC packet and writing it to the pre-allocated, coherent memory buffer, which is shared with the NIC. The hardware, i.e., the NIC and the memory sub-system, then handle the data transfer without any CPU intervention. This approach has the highest potential for networking systems of fine-grained workloads and latency-sensitive flows when aggressive batching is not desirable.

### 3.2 Design Principles of Dagger

Dagger adheres to the following design principles. First, the platform implements a hardware-based RPC protocol, i.e., the entire stack is offloaded to hardware. The software is only responsible for providing a zero-copy RPC API, alongside with the connection establishment and performance/statistics monitoring APIs. This improves networking performance by removing software and systems overheads from the critical path of RPC flows. In addition, this leaves more CPU resources available to applications, which is essential given the fine-grained and concurrent nature of cloud microservices.

Second, the design of Dagger is modular and fully reconfigurable. The Dagger RPC stack consists of a set of decoupled layers, as shown in Figure 3, and each layer can be programmed via either a soft register file (soft reconfiguration) or via FPGA-based reconfiguration of the bitstream (hard reconfiguration). In addition to programmable transports, Dagger also provides a reconfigurable CPU-NIC interface that allows fine-tuning the host-to-device communication model depending on the requirements and design patterns of target applications.

Third, Dagger leverages commercially-available, general-purpose memory interconnects instead of PCIe as the NIC I/O interface, which we showed to be more beneficial for fine-grained RPCs in Section 3.1. Dagger's current design is based on the Intel CCI-P interconnect built on top of the Intel UPI memory bus, as available in today's server-class processors. We focus on a readily available underlying system to implement Dagger on, to enable a near-future acceleration of cloud microservices without the need for taping-out custom chips and/or re-designing existing cloud infrastructures. The recent announcement of the Compute Express Link (CXL) [1], a new coherent interconnect dedicated specifically to peripheral devices, which is already supported in the newest generation of server processors and FPGAs shows the industry's interest in scaling memory interconnects out to peripherals. Given this trend, quantifying the performance and efficiency benefits of memory interconnects in performance-critical operations, such as networking, is essential to increasing their practicality.

### 3.3 Dagger System Overview

We implement Dagger using Intel Broadwell CPU/FPGA hybrid architecture; its top-level system diagram is shown in Figure 3. The host CPU runs the software part of the Dagger's RPC stack that includes the RPC API, transmission and receiving controllers, request buffers, etc., as shown in the CPU-NIC interface diagram of Figure 4. Dagger's software is running entirely in userspace and it communicates with the hardware via shared memory buffers and dedicated I/O regions over the CCI-P stack, with the latter encapsulating both UPI and PCIe interconnects.

The hardware component of the networking stack is built on top of the Intel OPAE HDK that defines two regions of programmable logic: the encrypted system IPs not exposed to users (blue bitstream), and the part of the FPGA available to users for implementation of the application logic (green bitstream). We implement Dagger in the green bitstream region labeled "NIC" in Figure 3. The NIC contains the RPC pipeline processing and transferring requests/responses, auxiliary components, such as the connection manager used to set-up, open, and close connections, the packet monitor for gathering networking statistics, and the soft reconfiguration unit for programming the register file of the RPC pipeline, enabling runtime configuration. The latter includes selecting the request batching size, the number of software buffers and their sizes, and the CCI-P messaging scheme, as defined in Section 3.4.

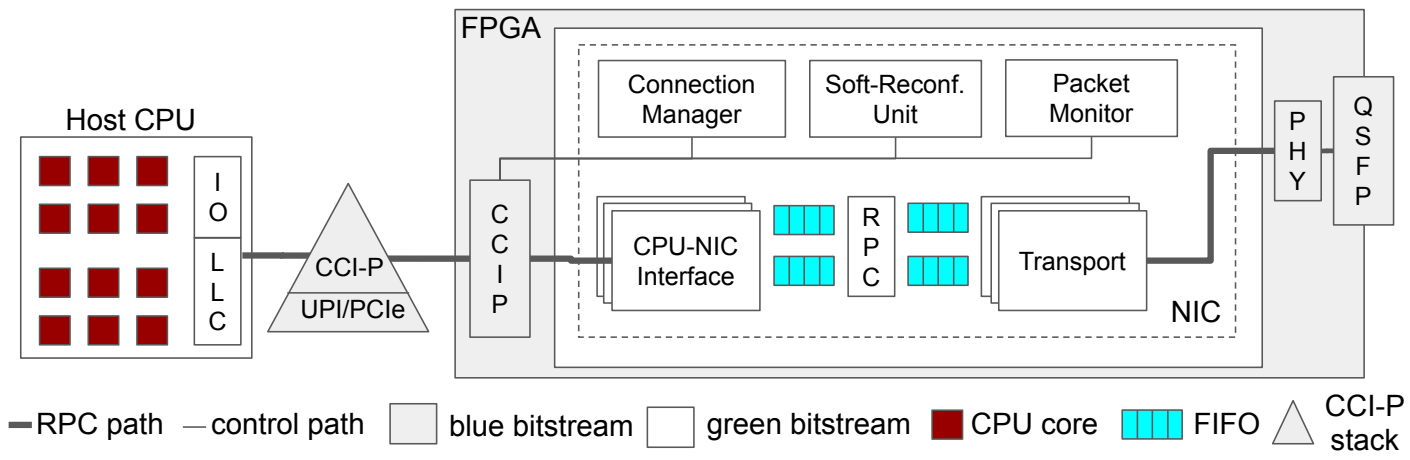

Fig. 3: Top-level diagram of the RPC acceleration fabric.

The RPC pipeline includes three main blocks: the CPU-NIC interface encapsulating CCI-P-based state machines for host-to-device data transfer, the RPC unit storing request metadata, and doing payload (de)serialization, and the transport serving as the protocol for transferring serialized RPCs. Stacked blocks in Figure 3 denote hard reconfigurable units. We make the transport layer reconfigurable to allow the flexibility of adjusting different protocol parameters depending on the requirements of the active applications and the characteristics of their network footprint. In our current Dagger prototype, the transport unit implements a version of the UDP protocol, however, any FPGA-based transport can be used. The CPU-NIC interface is also designed to be reconfigurable to adjust to a given running service. We discuss this in detail in the following section. Note that some parameters of the pipeline are soft-reconfigurable as explained earlier, allowing accelerating reconfiguration at runtime. However, since soft reconfiguration comes with logic overheads, the coarse-grained control decisions, such as the choice of the transport layer, are handled using hard reconfiguration.

### 3.4 Implementation of the CPU-NIC Interface

The CPU-NIC interface defines the exact scheme the NIC uses to communicate with application memory. The scheme includes (1) CCI-P physical/link medium and protocol messaging model, (2) RPC threading model, (3) network buffer provisioning, and (4) load balancing. An example of the Dagger's CPU-NIC interface is shown in Figure 4.

We first describe Dagger's CCI-P messaging scheme. The CCI-P interface supports three modes of host-to-device data transfers: over PCIe via MMIO, over PCIe via DMA, and over coherent memory interconnect (UPI). We implement all three modes in Dagger and select the employed scheme through hard reconfiguration; each scheme can be further fine-tuned via soft reconfiguration. In the MMIO mode, the processor sends RPC requests directly to the allocated I/O region of the FPGA via an AVX parallel store (_mm256_store_si256) with write combining disabled for lower latency. The optimization was initially proposed in [7] and is also known as WQE-by-MMIO in [11]. The DMA mode implements a version of the doorbell method including the doorbell batching optimization common in modern high-performance systems, such as Mellanox RDMA NICs. The third mode, UPI, is based on a memory interconnect and is the main focus of this work.

A typical memory interconnect defines a large set of different messaging schemes and memory consistency models that can be fine-tuned [1]. Unfortunately, since CCI-P is the first open-specification implementation of a NUMA interconnect on an FPGA, the current version of the CCI-P IP has limited functionality, and only allows accessing data via memory polling. In this mode, Dagger polls buffers shared with the CPU-running applications and explicitly tracks request updates via the dirty bit allocated in every entry of the buffer. To reduce CCI-P's bandwidth consumption by idle requests when no new data are available in the buffers, Dagger first polls its local cache, which is coherent with the processor's LLC, and relies on invalidation messages sent from the CPU to bring new data from the software buffers. When the request rate becomes high, as defined by a programmable threshold, Dagger switches by changing the caching policy of CCI-P requests to the direct polling of the CPU LLC, therefore providing higher bandwidth by eliminating invalidation transactions. Note that the restriction of data transfer to polling mode is simply an implementation artifact of the current CCI-P IP core and not a limitation of our design; the specification of the next generation peripheral memory interconnects (CXL) defines more optimized host-to-device transfers via invalidation snooping and direct in-device memory writes (Type 3 devices in [1]). We plan to explore these modes when the corresponding implementations become available. The device-to-host path, RX path in Figure 4, always uses CCI-P DMA since the write path is already optimized, and it contains only a single NIC-initiated DMA transaction per request in the critical path.

Dagger also defines three NIC buffer provisioning schemes: on connection basis, where every RPC connection gets the corresponding queue pair, on CPU core basis where all connections on the same core share TX and RX buffers, and single-buffer provisioning with only one TX and one RX buffer per NIC. The current version of Dagger only supports connection-based provisioning, as shown in Figure 4; the other schemes are currently under development.

Load balancing is closely related to buffer provisioning. Multiple load balancing schemes are possible when buffers are distributed across cores. For simplicity, the current design of Dagger supports fair round-robin load balancing where all RPC requests are evenly distributed across all CPU cores allocated on the server. We plan to explore more sophisticated load balancing schemes as part of future work.

Finally, Dagger implements both synchronous (blocking) and asynchronous (non-blocking) RPC threading models, and the choice of the exact model can be controlled by hard reconfiguration. The asynchronous model is shown in Figure 4, as it is the more complex of the two. The synchronous model is derived by simplifying the design with only the necessary components, for example, since synchronous RPCs block connections until the response is received, they do not require TX completion rings and bookkeeping transactions. As a result, synchronous RPCs save FPGA resources and further improve latency.

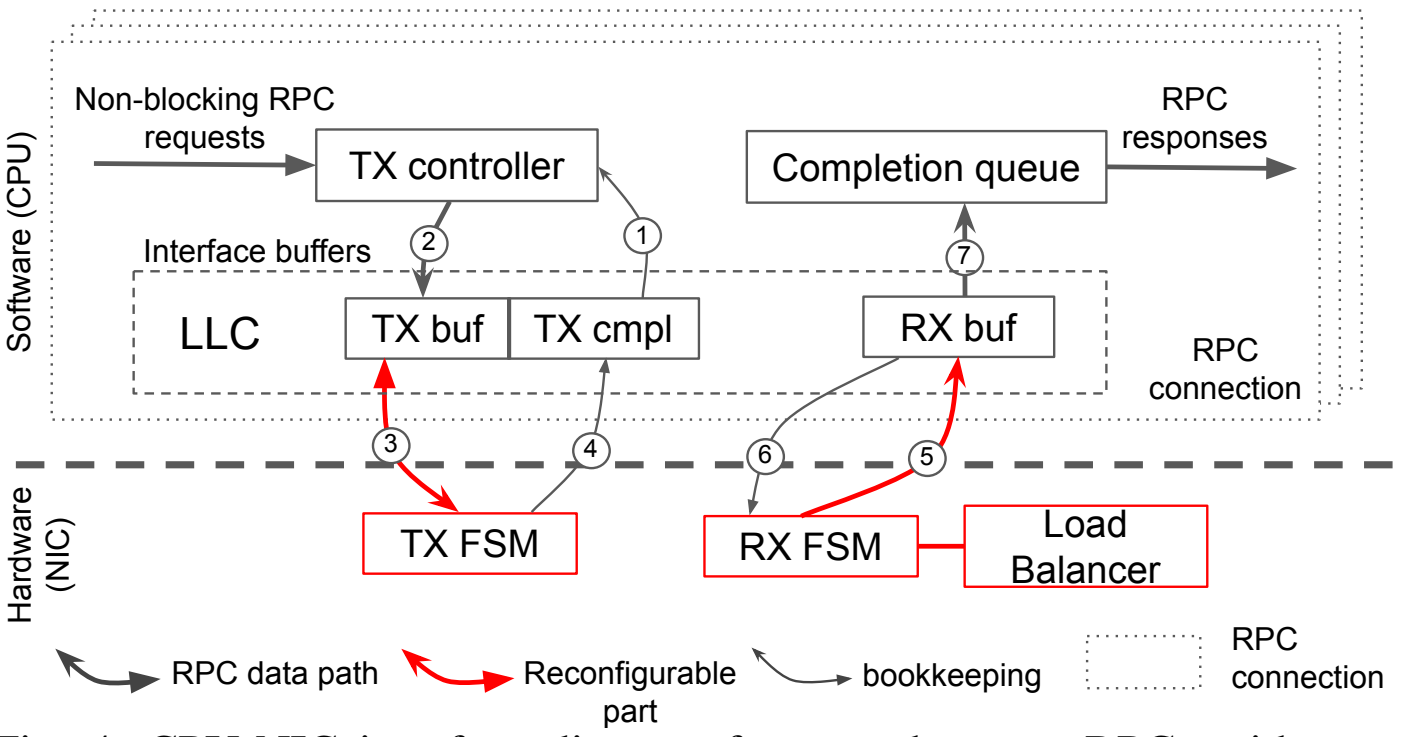

Fig. 4: CPU-NIC interface diagram for asynchronous RPCs with per-connection buffer provisioning. TX path: TX controller writes new RPC requests ② to a free entry in the TX buffer as read from the TX completion ring ①. The TX FSM on the FPGA fetches RPC objects from the TX buffer using one of the described CCI-P messaging schemes ③; it also does the bookkeeping ④ to release previously-fetched entries. RX path: the RX FSM puts newly-received RPC objects to the RX buffer and asynchronously fetches the next free entries by bookkeeping ⑥ both via CCI-P DMA. The RPC payload is then delivered to the completion queue by AVX-enhanced memcpy ⑦.

## 4 Preliminary Results

We implement Dagger on an Intel Broadwell CPU (Xeon E5-2600, 2.3GHz) integrated with an Arria 10 GX FPGA, as available on Intel vLab. Due to the hardware limitation of the vLab cluster, we cannot connect machines via physical networking; hence we conduct experiments on a single machine using the loop-back network on the FPGA. We place two identical Dagger NICs on one FPGA and give them fair round-robin access to the CCI-P bus. In our experiments, we run a concurrent P2P client-server application sending 64B echo RPCs. We analyze different host-to-device interface modes and compare the end-to-end performance of Dagger with related work that uses commercial NICs.

We first compare the performance of different CPU-NIC interfaces. Figure 5 shows that Dagger's single-core throughput with a simple doorbell ($B = 1$) method is 4.3 Mrps. This is already 65% higher than reported in FaSST's [12] single-core RPC throughput with the same doorbells, which indicates that a large fraction of Dagger's throughput gain comes from implementing a hardware-based RPC layer. Roughly the same performance of 4.2 Mrps is reached when Dagger is sending requests via MMIO, also showing that the throughput of doorbells is limited by the rate of PCIe MMIO transactions. The doorbell batching optimization pushes the throughput up to 7.9, 9.9, and 10.8 Mrps with B = 3, 7, and 11, respectively (same batching as in FaSST and eRPC for a fair comparison). When leveraging a memory interconnect as the CPU-NIC interface, Dagger achieves the throughput of 8.1 and 12.4 Mrps for CCI-P batching 1 and 4, which is 88% and 46% higher than doorbells with similar batching, shown in "UPI" bars of Figure 5). When pushing batching further to $B = 32$, RPC throughput approaches the results of the memory interconnect; however, it still saturates at 12 Mrps, which is slightly lower than the best reported performance over UPI.

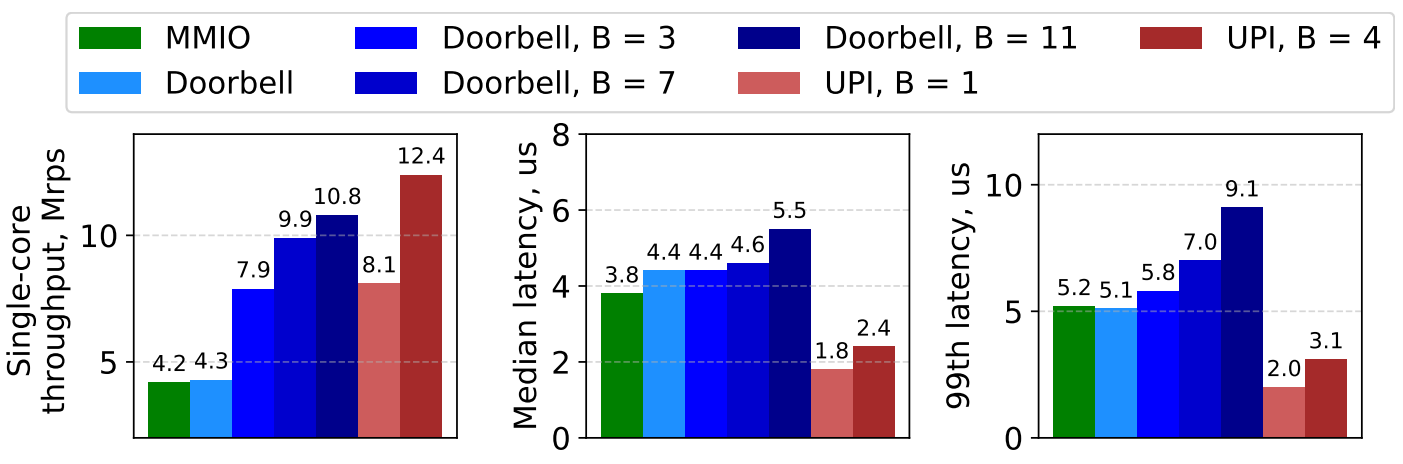

Fig. 5: Dagger's single-core throughput and latency (under 4 Mrps) comparison with different CPU-NIC interfaces; B denotes request batching.

Figure 5 also compares the median and tail latency of different CPU-NIC communication modes under the fixed load of 4 Mrps. Under all settings, Dagger outperforms both PCIe MMIO and all doorbell schemes, due to reducing the number of bus messages per RPC request, and replacing the slow MMIOs with faster memory transactions. In particular, the best recorded median latency of MMIOs is 3.8 us, while UPI allows latencies of 1.8 - 2.0 us when batching is disabled, and 2.4 - 3.1 us with B = 4. Since doorbells rely on MMIOs, all doorbell methods show higher median and tail latency than UPI. Moreover, doorbell batching has a negative impact on latency when the load is low. While this can be addressed with opportunistic batching [11], latency will never be lower than the latency of the simple doorbell; 4.4 us in our experiments.

Table 1 compares Dagger's best reported median round-trip time and single-core throughput with the best reported results presented by four related systems: IX [4], FaSST [12], eRPC [10], and NetDIMM [2]. We also show the objects being transferred, and the TOR networking delays assumed by each system for a fair comparison. Dagger significantly improves single-core RPC throughput for small requests compared to the four prior system, and it achieves similar latency to the highly-optimized eRPC system, and better than the DPDK-based IX, the RDMA-based FaSST, and the in-memory integrated NIC NetDIMM.

Table 1: Median round trip time and throughput of asynchronous single-core RPCs compared to related work.

| | **IX** | **FaSST** | **eRPC** | **Net-DIMM** | **Dagger** |
|---|---|---|---|---|---|
| **Objects** | 64B msgs | 48B RPC | 32B RPC | 64B msgs | 64B RPC |
| **TOR delay** | N/A | 0.3 us | 0.3 us | 0.1 us | 0.3 us |
| **RTT, us** | 11.4 | 2.8 | 2.3 | 2.2 | 2.1 |
| **Thr., Mrps** | 1.5 | 4.8 | 4.96 | N/A | 12.4 |

Figure 6 (left) shows the latency-throughput curves across loads. Since we run a simple echo microbenchmark, the system immediately blocks the caller thread when throughput is saturated, shown with vertical dotted lines. The latency remains stable across the entire load range. The figure also shows how UPI batching might affect the system performance: B = 4 allows high throughput, however, it also increases request latency for low QPS. Since Dagger can adjust the batch size of the interconnect on-the-fly via soft reconfiguration, it can dynamically change batching depending on the current load, as shown by the green dashed line.

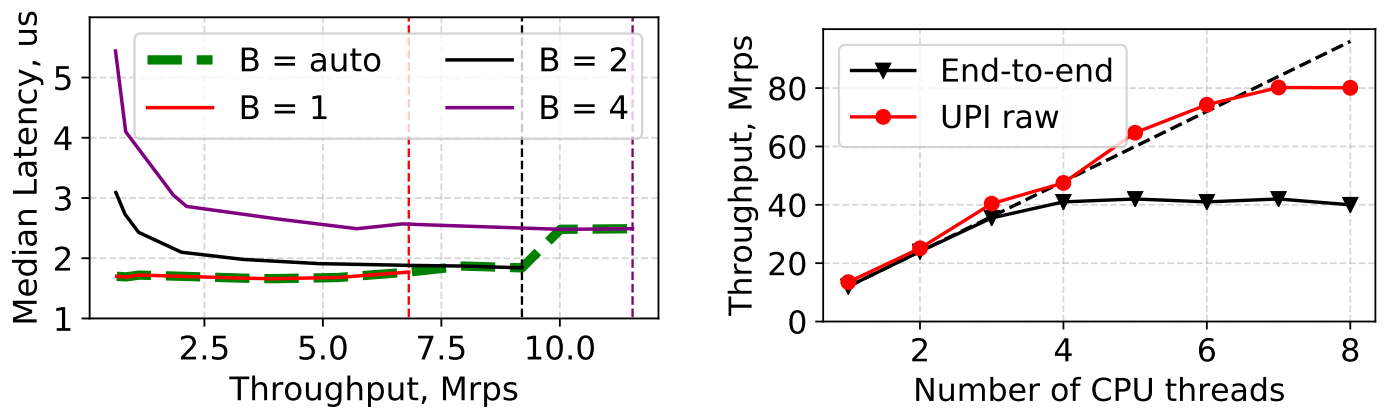

Fig. 6: (Left) Latency-Throughput curves for single-core asynchronous round-trip 64B RPCs; B denotes batching, dotted lines show the saturation point. (Right) Multi-core scalability of sending 64B requests. The black line shows end-to-end RPC throughput, the dashed line denotes estimated linear scalability, and the red line shows the results of UPI bus scalability with raw idle requests.

Figure 6 (right) shows the throughput scalability of Dagger with the number of physical threads (logical cores). Dagger achieves 42 Mrps for end-to-end client-server median throughput with 4 threads (2 cores). This result is ∼23% better than the best RDMA-based solution, FaSST [12], and 3.5× better than DPDK-based IX [4]. Since we run both the client and server on the same CPU, 42 Mrps effectively translates to 84 Mrps handled by the processor. Dagger's throughput scales linearly up to 4 threads and remains flat at 40-42 Mrps, showing that the current bottleneck is not CPU. The UPI bus is also not saturated: 84 Mrps of 64B requests is 7.74× smaller than the 41.6 GB/s of available UPI bandwidth. Similarly, the NIC running at 400 MHz is capable of processing up to 200 Mrps, and is therefore far from saturation. We find that the current bottleneck limiting throughput to 84 Mrps is the implementation of the UPI end-point on the FPGA that we, unfortunately, have no control over. To confirm this, we run a simple benchmark sending raw 64B memory requests over CCI-P and measure the achieved throughput. As seen from Figure 6 (right, red line), the raw throughput scales linearly up to 80 Mrps and stays flat when more cores are added. The main reason behind its poor scalability is that the current version of the OPAE HDK, implements the memory interconnect as a soft IP core, i.e., using FPGA LUT resources. The next generation of FPGAs (Intel Agile) implements a coherent memory interconnect (CXL) in hardware as an integrated ASIC. This, together with the more efficient host-to-device communication modes discussed in Section 3.4) will enable a faster CPU-NIC interface, and higher throughput, lower latency, and better scalability.

## 5 Conclusion

We presented Dagger, our early work on hardware-accelerated RPCs for cloud microservices, using a closely-coupled near-memory FPGA connected over a memory interconnect to the host CPU. Dagger significantly increases the per-core RPC throughput, while achieving state-of-the-art round-trip latency in comparison with existing RPC frameworks using commodity PCIe-attached NICs. Our prototype outperforms existing user-space networking- and RDMA-based solutions based on peripheral networking devices, and, additionally provides transparent NIC I/O at the hardware level. Overall, we show that closely-coupling CPUs and FPGAs can provide efficient and programmable networking that drastically improves performance for interactive, multi-tier cloud services.

## Acknowledgements

We sincerely thank the anonymous reviewers for their feedback on earlier versions of this manuscript. This work was in part supported by an NSF CAREER Award CCF-1846046, NSF grant NeTS CSR-1704742, NSF/Intel CAPA grant CCF-1723715, a Sloan Research Fellowship, a Microsoft Research Fellowship, and a John and Norma Balen Sesquisentennial Faculty Fellowship.

## References

[1] "Compute express link (CXL) specification," accessed May, 2020, https://www.computeexpresslink.org/.
[2] M. Alian and N. S. Kim, "NetDIMM: Low-latency near-memory network interface architecture," *Int'l Symp. on Microarchitecture (MICRO)*, 2019.
[3] M. T. Arashloo, A. Lavrov *et al.*, "Enabling programmable transport protocols in high-speed NICs," *USENIX Symp. on Networked Systems Design and Implementation (NSDI)*, 2020.
[4] A. Belay, G. Prekas *et al.*, "IX: A protected dataplane operating system for high throughput and low latency," *USENIX Symp. on Operating Systems Design and Implementation (OSDI)*.
[5] A. Dragojević, D. Narayanan *et al.*, "FaRM: Fast remote memory," *USENIX Symp. on Networked Systems Design and Implementation (NSDI)*, 2014.
[6] H. Eran, L. Zeno *et al.*, "NICA: An infrastructure for inline acceleration of network applications," *USENIX Annual Technical Conf. (ATC)*, Jul. 2019.
[7] M. Flajslik and M. Rosenblum, "Network interface design for low latency request-response protocols," *USENIX Annual Technical Conf. (ATC)*, 2013.
[8] Y. Gan, Y. Zhang *et al.*, "An open-source benchmark suite for microservices and their hardware-software implications for cloud and edge systems," *ASPLOS*, 2019.
[9] Q. Huang, K. Birman *et al.*, "An analysis of facebook photo caching," *ACM Symp. on Operating Systems Principles (SOSP)*, 2013.
[10] A. Kalia, M. Kaminsky *et al.*, "Datacenter RPCs can be general and fast," *USENIX NSDI*, 2019.
[11] A. Kalia, M. Kaminsky *et al.*, "Design guidelines for high performance RDMA systems," *USENIX Annual Technical Conf. (ATC)*, 2016.
[12] A. Kalia, M. Kaminsky *et al.*, "FaSST: Fast, scalable and simple distributed transactions with two-sided (RDMA) datagram RPCs," *USENIX Symp. on Operating Systems Design and Implementation (OSDI)*, 2016.
[13] R. Neugebauer, G. Antichi *et al.*, "Understanding PCIe performance for end host networking," *ACM Special Interest Group on Data Communication (SIGCOMM)*, 2018.
[14] S. Novakovic, A. Daglis *et al.*, "Scale-out NUMA," *Int'l Conf. on Architectural Support for Programming Languages and Operating Systems (ASPLOS)*, 2014.
[15] P. M. Phothilimthana, M. Liu *et al.*, "Floem: A programming system for NIC-accelerated network applications," *Symposium on Operating Systems Design and Implementation (OSDI)*, 2018.
[16] A. Pourhabibi, S. Gupta *et al.*, "Optimus prime: Accelerating data transformation in servers," *Int'l Conf. on Architectural Support for Programming Languages and Operating Systems (ASPLOS)*, 2020.
[17] D. Sidler, Z. István *et al.*, "Low-latency TCP/IP stack for data center applications," *Int'l Conf. on Field Programmable Logic and Applications (FPL)*, 2016.
[18] A. Sriraman, A. Dhanotia *et al.*, "SoftSKU: Optimizing server architectures for microservice diversity at scale," *Int'l Symp. on Computer Architecture (ISCA)*, 2019.
[19] A. Sriraman and T. F. Wenisch, "µsuite: A benchmark suite for microservices," *Int'l Symp. on Workload Characterization (IISWC)*, 2018.
[20] A. Sriraman and T. F. Wenisch, "µtune: Auto-tuned threading for OLDI microservices," *USENIX Symp. on Operating Systems Design and Implementation (OSDI)*, 2018.
[21] M. Sutherland, S. Gupta *et al.*, "The NeBuLa RPC-optimized architecture," *Int'l Symp. on Computer Architecture (ISCA)*, 2020.
[22] Y. Xu, E. Frachtenberg *et al.*, "Characterizing facebook's memcached workload," *IEEE Internet Computing*, 2014.